*Communication*

# Photo-Writable Sulfide Glasses Used to Fabricate Core-Clad Fiber Doped with Pr$^{3+}$ for Mid-IR Luminescence

Julie Carcreff [1], Virginie Nazabal [1], Johann Troles [1], Catherine Boussard-Plédel [1], Pascal Masselin [2], Florent Starecki [3], Alain Braud [3], Patrice Camy [3] and David Le Coq [1,*]

1 CNRS, ISCR-UMR 6226, Université Rennes, 35000 Rennes, France
2 Laboratoire de Physico-Chimie de l'Atmosphère, Université du Littoral-Côte d'Opale, 59140 Dunkerque, France
3 CIMAP, UMR 6252 CEA-CNRS-ENSICAEN, Université de Caen Normandie, 14050 Caen, France
* Correspondence: david.lecoq@univ-rennes1.fr

**Abstract:** With the ultimate goal of developing rare-earth doped chalcogenide fiber fabrication for sensing, amplification, and laser applications, a core/clad germanium-gallium sulfide fiber doped with Pr$^{3+}$ has been fabricated. The compositions of the core and the clad were selected to ensure the positive ∆n by adding CdI$_2$ and CsCl, respectively, in the GeS$_2$-Ga$_2$S$_3$ matrix. The choice of these compositions was also justified from experimental parameters, including characteristic temperatures and viscosity. Moreover, the permanent photo writability of the sulfide glass family by a femtosecond laser is investigated from the perspective of Bragg grating photo-inscription. Structural investigations by Raman spectroscopy are presented and the effect of the Pr$^{3+}$ rare-earth ion on the structure is underlined. Finally, the emission of the step-index fiber, made by the rod-in-tube technique between 3.1 µm and 5.5 µm (by pumping at 1.55 µm), is demonstrated.

## 1. Introduction

The development of light sources in the mid-infrared range (MIR) has attracted increasing interest in the field of photonics over the last decade. As a matter of fact, they can find many areas of applications such as: (i) remote sensing for gases and chemical pollutants; (ii) medicine (since human tissues can be treated with MIR lasers); and (iii) national security and defense using MIR countermeasures. However, many of these applications in the MIR require compact and robust sources with an affordable price, high efficiency and output power, and good output beam quality.

In this context, up to now, only a few laser oscillations were observed in chalcogenide glass (ChG) fibers [1]. Over the last 25 years, several research teams have tried to obtain a laser action in the 3–5 µm spectral range. Nevertheless, the complexity and the difficulty of this fabrication of rare earth doped chalcogenide waveguides to obtain a laser gain are still present despite some very recent achievements [1]. One of the main causes lies firstly in the optical losses of a double index fiber at the excitation and the laser emission wavelengths. Second, the extrinsic non-radiative relaxations due to the presence of impurities, respectively, weaken the RE emission in the MIR. Recently, numerical simulations have provided new hopes and have been proposed as promising cascading lasing systems for Dy$^{3+}$ doped chalcogenide fiber at 4.2 µm [2]. Subsequently, a MIR fiber laser based on Dy$^{3+}$ doped sulfide glass was proposed [3]. Other papers relate the detailed theoretical study of laser amplification in Tb-doped chalcogenide fiber laser operating at 5.25 µm [4] or Pr-doped chalcogenide fibers pumped at 1.55 or 2 µm [5]. Analogously, a promising MIR laser numerical modeling was also developed in the case of Pr$^{3+}$ doped ChG fiber [6,7].

Up to now, the most powerful fiber lasers realized in the MIR spectral range use ZBLAN fibers doped with RE ions [8]. The longest operating wavelength achieved is below 4.5 µm [9] due to the multiphonon relaxation in fluoride glasses that quenches the luminescence from higher energetic states. ChG have a lower phonon energy to achieve laser emission at longer wavelengths as demonstrated by calculations [10] and also possess a high refractive index, which is beneficial to obtain higher absorption and emission cross-sections in RE-doped glasses. Indeed, it should be noticed that laser emission up to 7.2 µm has been demonstrated in RE-doped halogenide crystal [11], but their high hygroscopicity makes them practically unusable (whereas this is not the case for ChG). The MIR luminescence of different RE in the ChG bulks or fibers for wavelengths beyond 4 µm has already been shown [12] and very recently a first MIR fiber lasing has been reported in the 5.1–5.2 µm range [13]. Nevertheless, up to now, the fabrication of a convenient MIR fiber laser in which the laser cavity will be directly embedded in the fiber core is not demonstrated yet. This last point requires for example the presence of Bragg mirrors (BM) inside the fiber core that could be inscribed by femtosecond laser writing. In other words, the composition of the ChG must allow simultaneously: (i) a femtosecond writing of the BM at the wavelength of inscription; and (ii) a high level of RE doping. The Ge-Ga-S glasses, which are transparent in the visible range, are suitable for laser writing at 800 nm [14]. The presence of gallium also facilitates the incorporation and the dispersion of RE ions by limiting the cluster formation [15]. The addition of metal halides in the glass composition facilitates the synthesis process and allows an adjustment of glass properties necessary to define the core and clad formulation. In this paper, we first investigate the effects on the thermal properties of CsCl and $CdI_2$ addition in the glass matrix composition of $[GeS_2]_{0.80}$-$[Ga_2S_3]_{0.20}$, and their behavior in terms of refractive index modification under femtosecond laser irradiations at 800 nm. This work is devoted to Raman structural analyses of the chosen clad and $Pr^{3+}$-doped core compositions and to evaluating its emission performance.

## 2. Materials and Methods

*2.1. Syntheses*

The glasses from the systems $([GeS_2]_{0.80}$-$[Ga_2S_3]_{0.20})_{100-x}$-$(CdI_2)_x$ and $([GeS_2]_{0.80}$-$[Ga_2S_3]_{0.20})_{100-x}$-$(CsCl)_x$ were synthesized from Ge (Umicore, 5 N), Ga (Alfa Aesar, 7 N), and S (Strem Chemicals, 5 N), as raw elements and from $CdI_2$ (Alfa Aesar, 5 N) or CsCl (Alfa Aesar, 5N) compounds. The rare earth ions were added as $Pr_2S_3$ for the core glass doped with 1000 ppmw of $Pr^{3+}$. The syntheses of the glasses were implemented by the melt-quenching method in a vacuum sealed silica set-up. The sulfur was distilled to remove traces of carbon, hydrogen, sulfide oxide, and water [16]. After homogenization of the molten liquid at 1023 K for several hours, the rods were obtained by quenching in water at room temperature and next an annealing at a temperature below the glass transition temperature (Tg), namely Tg-15 K [17]. The 7- or 10-mm diameter glass rods were then polished before their uses for the elaboration of fibers.

*2.2. Thermal Characterizations*

The glass transition temperatures (Tg) were determined by differential scanning calorimetric analyses on a DSC Q20 (Thermal Analysis Instrument). A sample mass of 5 to 10 mg was taken and introduced into an aluminum crucible which was then sealed and placed in the DSC furnace. Next, a temperature ramp of 10 K per minute was applied up to a maximum temperature of 773 K. For all compositions, no crystallization phenomenon was observable on the DSC traces.

The viscosity was measured by the parallel plate method on a Rheotronic® viscometer (Theta industries). A sample whose sides have been polished was placed between two plates. Then, a probe is placed on top with a constant pressure of 200 g. The glass is heated at a rate of 2 K·min$^{-1}$ up to 823 K.

The thermal expansion coefficient α of the glasses was evaluated by thermomechanical analysis on a TMA 2940 dilatometer from the TA Instrument using Equation (1):

$$\alpha = \frac{\Delta L_s}{L_0(T_{max} - T_{amb})} \quad (1)$$

with $\Delta L_s$ is the sample length variation (mm), $L_0$ is the sample length (mm) at room temperature, $T_{amb}$ (K), and $T_{max}$ (K) is the maximal temperature of measurement. The measurement was performed up to Tg-10 K with a temperature rise rate of 2 K·min$^{-1}$ under a constant pressure of 0.1 N.

*2.3. Structural Characterizations*

Raman scattering spectra are recorded for the two bulk glass compositions (([GeS$_2$]$_{0.80}$-[Ga$_2$S$_3$]$_{0.20}$)$_{90}$-(CsCl)$_{10}$ and [GeS$_2$]$_{0.80}$-[Ga$_2$S$_3$]$_{0.20}$)$_{90}$-(CdI$_2$)$_{10}$:1000 ppm Pr$^{3+}$ used as cladding and core glasses, respectively. A Horiba Scientific LabRAM HR Evolution spectrometer was used with a 785 nm laser diode as excitation source.

*2.4. Step-Index Fiber Elaboration*

The preforms were prepared by the rod-in-tube method. The composition (([GeS$_2$]$_{0.80}$-[Ga$_2$S$_3$]$_{0.20}$)$_{90}$-(CsCl)$_{10}$ that was chosen for the clad was pierced and a rod of composition (([GeS$_2$]$_{0.80}$-[Ga$_2$S$_3$]$_{0.20}$)$_{90}$-(CdI$_2$)$_{10}$:1000 ppm Pr$^{3+}$ was prepared for the core. In a recent paper, we have already determined the refractive indexes (n) of the glasses of these series. At similar x, the CdI$_2$-based glasses possess a higher n [16]. In the chosen core and clad compositions, they are 2.074 and 2.039, respectively. The fibering process used to elaborate the step-index fiber has also been previously described in [16]. As a result, the preform was drawn into a 350 µm diameter core/cladding fiber with a core diameter of 60 µm.

*2.5. Optical Characterizations*

2.5.1. Femtosecond Laser Δn Modification in the Glass Bulks

Usually, refractive index variations induced by femtosecond laser photo-inscription are carried out by translating the sample at constant speed, either parallel to the beam (longitudinal geometry) or perpendicular (transverse geometry). However, another methodology was applied in order to be consistent with the successful process used for the realization of optical components [18,19]. The inscriptions were made in a longitudinal geometry. In the first step, the beam was focused in the sample using a f = 50 mm lens and this one was irradiated during a time τ, by a pulse train of duration 250 fs with a repetition rate of 250 kHz, inducing the appearance of Δn. The longitudinal displacement Δz is about 10 µm. This value is chosen in order to obtain a good overlapping between the steps and then a good homogeneity of the resulting Δn channel. These sequences were repeated for the entire length of the sample. Although the process is discontinuous, the resulting index variation is homogeneous.

The inscriptions were made in the bulk of the glass in pieces of dimensions 7 × 7 × 40 mm$^3$. The samples were cut and polished into a 600 µm thickness plate before the evaluation of Δn. Measurements of photo-induced index changes were performed by quantitative phase microscopy imaging [20] under a microscope with an X50 objective. On the other hand, we can notice that the longitudinal inscription geometry implies an axial symmetry of Δn. Also, we can apply an inverse Abel transformation to the previous phase measurements in order to determine the profile of the index variation Δn(r) [21]. We note that several measurements along the Δn channel were made in order to verify the homogeneity of the results of the writing procedure.

2.5.2. Fluorescence Measurements

Fluorescence spectra were recorded using a diode emitting at 1.55 µm (Q-Photonics QSM 1550-3, Ann Arbor, USA) with incident power of 1030 mW, a liquid nitrogen cooled InSb detector adapted diffraction gratings for each wavelength domain and long-pass filter (Spectrogon LP-2500, Stockholm, Sweden) to remove any parasitic higher order contribution from the transmitted pump.

## 3. Results and Discussion

### 3.1. Thermal Properties

The evolutions of Tg as a function of the metal halide CsCl or $CdI_2$ molar contents (*x*) are exhibited in Figure 1 for the glasses belonging to the two series ($[GeS_2]_{0.80}$-$[Ga_2S_3]_{0.20})_{100-x}$-$(CdI_2)_x$ and ($[GeS_2]_{0.80}$-$[Ga_2S_3]_{0.20})_{100-x}$-$(CsCl)_x$ for $0 \leq x \leq 20$. In both cases, a decrease of Tg is observed with the increase of *x*. This feature has mainly to be correlated to a dereticulation of the glass network due to the addition of halogens, which break the Ge-S or Ga-S bonds. More details on the structure of these glasses are given in the Section 3.3. Also, at constant metal halide molar content, the Tg difference between the CsCl- and $CdI_2$-based glasses are never higher than 15 K. For example, at *x* = 10%, Tg are 668 K and 653 K for $[GeS_2]_{0.80}$-$[Ga_2S_3]_{0.20})_{90}$-$(CsCl)_{10}$ and $[GeS_2]_{0.80}$-$[Ga_2S_3]_{0.20})_{90}$-$(CdI_2)_{10}$, respectively.

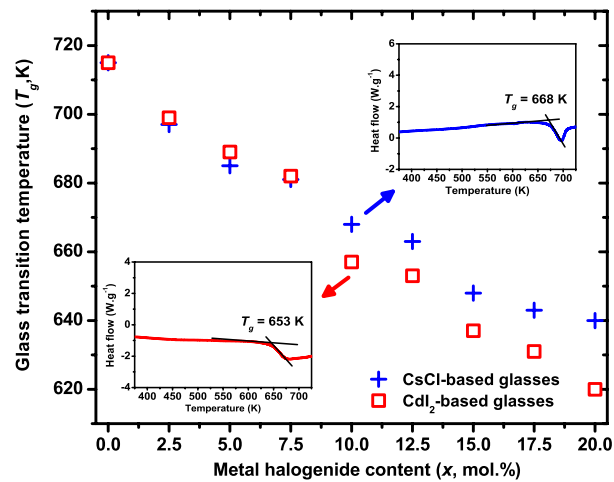

**Figure 1.** Evolution of $T_g$ as a function the metal halide CsCl (blue crosses) or $CdI_2$ (red squares) content. Inset correspond to the DSC traces of glasses with 10 mol. % of CsCl (blue line) or $CdI_2$ (red line).

In addition, to shape a glass, the viscoplastic regime must show a viscosity between $10^8$ and $10^3$ Pa·s and its temperature range must be between Tg and the onset crystallization temperature Tx. In general, the fiber drawing for a chalcogenide glass is facilitated when the viscosity is below $10^6$ Pa·s [22]. The temperature at which this level is reached gives an indication of the fiber drawing temperature. Figure 2a exhibits the evolution of the temperature for a viscosity of $10^6$ Pa·s as a function of the molar content in metal halide. For glasses belonging to the two investigated series, this temperature monotonously decreases with *x* (and more pronouncedly for the $CdI_2$-based glasses). This last observation has to be correlated to the halogen content, which is higher for $CdI_2$- than for CsCl-based glasses at equal metal halide content. This trend is in good agreement with the decreasing $T_g$ evolution as a function of *x* that is more important in the $CdI_2$-based glasses (Figure 1). Moreover, Figure 2b shows that the thermal expansion coefficient ($\alpha$) increases with the metal halide content for the series based on CsCl, while it remains almost constant for the series containing $CdI_2$. Indeed, $\alpha$ grows from $10 \times 10^{-6}$ K$^{-1}$ up to $26 \times 10^{-6}$ K$^{-1}$ for *x* equal to 5 and 20 in CsCl-based glasses, respectively, whereas it stagnates around $9.5 \times 10^{-6}$ K$^{-1}$ for $CdI_2$-based glasses. These features can be attributed to the size of the cations since, according to Pauling radii, the ionic radii of Cs$^+$ and Cd$^{2+}$ are 1.69 Å and 0.97 Å, respectively.

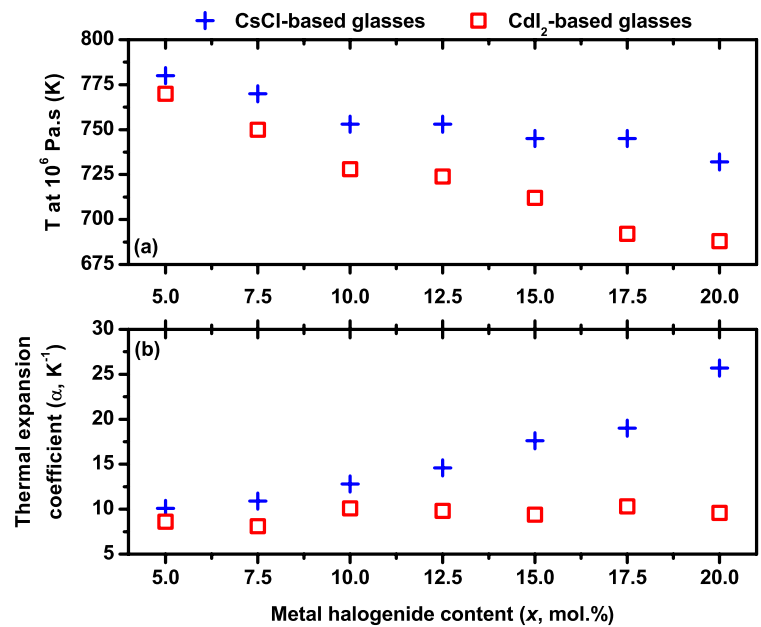

**Figure 2.** (**a**) Evolution of the measured temperature at a viscosity of $10^6$ Pa·s as a function the metal halide CsCl (blue crosses) or CdI$_2$ (red squares) content. (**b**) Evolution of the thermal expansion coefficient as a function the metal halide CsCl (blue crosses) or CdI$_2$ (red squares) content.

The data in Table 1 allow us to compare the thermal characteristics of the composition of the fiber core undoped and Pr$^{3+}$-doped. The addition of praseodymium does not significantly influence the thermal properties, meaning that the behavior of the doped glass during the drawing process will probably remain unchanged.

**Table 1.** Thermal characteristics of glasses corresponding to the undoped and Pr$^{3+}$-doped cores, [GeS$_2$]$_{0.80}$-[Ga$_2$S$_3$]$_{0.20}$)$_{90}$-(CdI$_2$)$_{10}$ and [GeS$_2$]$_{0.80}$-[Ga$_2$S$_3$]$_{0.20}$)$_{90}$-(CdI$_2$)$_{10}$:1000 ppmw Pr$^{3+}$, respectively.

| Glass Composition | Tg (K) ± 2 K | Tx (K) ± 2 K | ΔT (K) ± 4 K | α (10$^{-6}$ K$^{-1}$) ± 0.5 × 10$^{-6}$ K$^{-1}$ | T at 10$^6$ Pa·s ± 2 K |
|---|---|---|---|---|---|
| Undoped core | 653 | >773 | >120 | 10.1 | 728 |
| Pr$^{3+}$-doped core | 654 | >773 | >119 | 10.5 | 730 |

*3.2. Refractive Index Modification by Femtosecond Laser*

In our study, two types of experiments are presented. In the first one, shown in Figure 3a, the average beam power is fixed (P = 20 mW) and the duration of the pulse burst τ varies. In the second case, Figure 3b, τ is fixed at 100 ms and the study focuses on the influence of the incident power. The results are quite similar in both series for the 2 experiments. First of all, when P is fixed at 20 mW, the generated Δn reach 6.10$^{-3}$ for a duration pulse higher than 125 ms. Next, for τ = 100 ms, a threshold appears around 10 mW. Below this power the Δn is negligible, whilst above 15 mW it reaches 6.10$^{-3}$. Finally, the previous works published elsewhere [14,19] and the present studies of refractive index modification by femtosecond laser allow us to positively consider this technique for inscribing BM inside the core of the fiber.

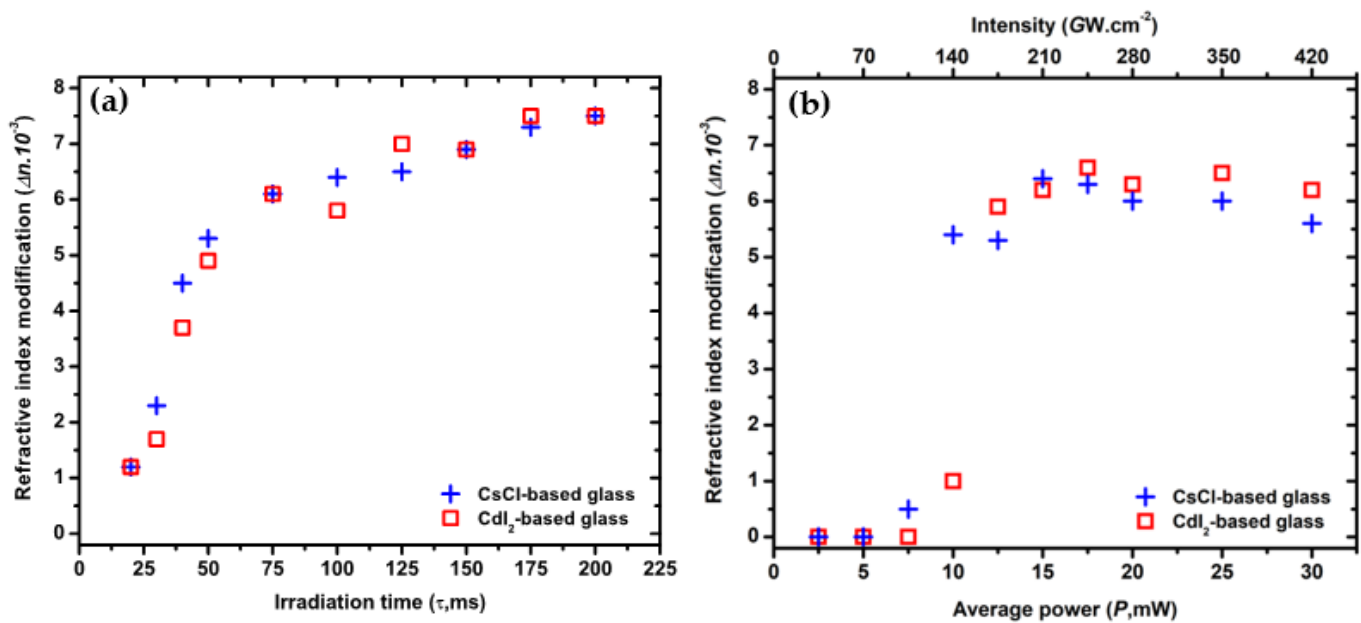

**Figure 3.** Evolution of the refractive index modification (Δn) as a function of (**a**) the irradiation time (τ) at 20 mW and (**b**) the average power (P) and the corresponding intensity with τ = 100 ms in the selected glasses for the step-index fiber (CsCl—(blue cross) and CdI$_2$—(red square) based glasses.

*3.3. Glass Structure*

The Raman spectra of [GeS$_2$]$_{0.80}$-[Ga$_2$S$_3$]$_{0.20}$)$_{90}$-(CsCl)$_{10}$ and [GeS$_2$]$_{0.80}$-[Ga$_2$S$_3$]$_{0.20}$)$_{90}$-(CdI$_2$)$_{10}$:1000 ppmw Pr$^{3+}$ bulk glasses are dominated, as can be seen in Figure 4, by the presence of a band at 340 cm$^{-1}$ and a shoulder observed at 370 cm$^{-1}$. These band and shoulder correspond to the symmetric $\nu_1(A_1)$ stretching modes of the [GeS$_4$] tetrahedra[23] bound by their corners and to the companion mode of the $\nu_1$ symmetric streching modes related to vibrations of the [GeS$_4$] tetrahedra linked by their edges. A band at 430 cm$^{-1}$ is assigned to vibrations of S$_3$Ge-S-GeS$_3$ units where the tetrahedra are connected by their corners and the asymmetric streching mode of [GeS$_4$] tetrahedra is expected to present a moderate contribution at 385–405 cm$^{-1}$. A broadening of the dominant peak is associated with the presence of gallium in Ge-based glass, which can be attributed to the symmetric stretching mode of [GaS$_4$] tetrahedra (around 320–350 cm$^{-1}$), usually accompanied by S$_3$Ga-S-GaS$_3$ units around 360–390 cm$^{-1}$ [24–28]. In case of a deficit in sulfur in glass composition, a triscluster of [GaS$_4$] sharing one S tricoordinated as in the case of α-Ga$_2$S$_3$ crystal, was also proposed and calculated from the density functional theory (DFT) simulation at 315 and 397 cm$^{-1}$ or 325 cm$^{-1}$ without clear attribution in experimental Raman spectrum [29,30]. The possible presence of an edge-sharing (ES) [GaS$_4$] unit was proposed to present vibration at 240 cm$^{-1}$ [29] or 270 cm$^{-1}$ of two edge-shared tetrahedra [31]. The band located at 474 cm$^{-1}$ attributed to the stretching vibrational modes of homopolar S-S bonds, which can form dimers, small chains, or S$_8$ rings is only very weakly present in ([GeS$_2$]$_{0.80}$-[Ga$_2$S$_3$]$_{0.20}$)$_{90}$-(CsCl)$_{10}$ and ([GeS$_2$]$_{0.80}$-[Ga$_2$S$_3$]$_{0.20}$)$_{90}$-(CdI$_2$)$_{10}$:1000 ppm Pr$^{3+}$ from Pr$_2$Se$_3$. The presence of CsCl in Ge-Ga-S seems to have a not very pronounced impact on the whole of the spectrum, probably causing a broadening of bands with really spread contributions over the whole of the spectrum and generally related to an insertion of chlorine in the entities tetrahedral via a substitution of S by Cl. On the other hand, we note a retively noticeable difference on the bands between 240 and 270 cm$^{-1}$ between the CsCl and CdI$_2$ glasses. A small band is observed at 270 cm$^{-1}$ for ([GeS$_2$]$_{0.80}$-[Ga$_2$S$_3$]$_{0.20}$)$_{90}$-(CsCl)$_{10}$ and additionaly, a weak shoulder at 258 cm$^{-1}$ and a medium band at 240 cm$^{-1}$ for ([GeS$_2$]$_{0.80}$-[Ga$_2$S$_3$]$_{0.20}$)$_{90}$-(CdI$_2$)$_{10}$. In this spectral range, the vibration mode of S$_3$Ga-Ge(Ga)S$_3$ can be expected around 268 cm$^{-1}$, for which it is difficult to distinguish Ge-Ga bonds from Ga-Ga bonds due to their very close atomic weight. The band at 240 cm$^{-1}$ is already observed as a clear shoulder of the 270 cm$^{-1}$ band in Na$_2$S-Ga$_2$S$_3$-CsCl and GeS$_2$-Ga$_2$S$_3$ glasses when Ga$_2$S$_3$ increased and could be related to vibration of two edge-shared [GaS$_4$] tetrahedra or [Ga$_y$-GaS$_{4-y}$]. The shoulder may be associated with the vibrational modes of Ge-Ge homopolar bonds (likely at 258 cm$^{-1}$) existing in the S$_3$Ge-GeS$_3$ ethane-type units or [Ge$_y$-GeS$_{4-y}$] for sulfur-deficient Ge-based glasses. It was also reported in GeS$_2$- CdI$_2$ and GeS$_2$-Ga$_2$S$_3$-CdI$_2$ glasses the presence of a double band at 240 cm$^{-1}$ and 258 cm$^{-1}$ which clearly increases with increasing CdI$_2$. This could be related to the presence of two edge-shared [GaS$_4$] tetrahedra or [Ga$_y$-GaS$_{4-y}$] and S$_3$Ge-GeS$_3$ in greater proportion but it is also proposed that the presence of both [GaS$_3$I] and [GeS$_3$I] are susceptible to the present contribution.

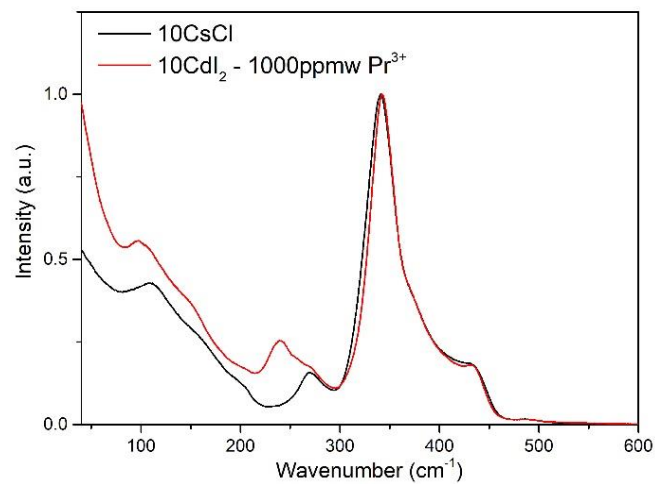

**Figure 4.** Raman spectra of the glasses ([GeS$_2$]$_{0.80}$-[Ga$_2$S$_3$]$_{0.20}$)$_{90}$-(CsCl)$_{10}$ and ([GeS$_2$]$_{0.80}$-[Ga$_2$S$_3$]$_{0.20}$)$_{90}$-(CdI$_2$)$_{10}$:1000 ppmw Pr$^{3+}$ recorded at 785 nm.

*3.4. MIR Emission of Core-Clad Sulfide Fiber*

First of all, the optical losses of the double index fiber ([GeS$_2$]$_{0.80}$-[Ga$_2$S$_3$]$_{0.20}$)$_{90}$-(CsCl)$_{10}$ and ([GeS$_2$]$_{0.80}$-[Ga$_2$S$_3$]$_{0.20}$)$_{90}$-(CdI$_2$)$_{10}$:1000 ppmw Pr$^{3+}$ have been estimated by the cut-back method. Typically, these optical losses are about a few tens of dB.m$^{-1}$ between 2.5 and 8 µm. Even if this characteristic is not satisfactory, the emission spectrum of the core of the doped step-index fiber was still measured. A broad emission band is, as expected, observed in the 3.2 to 5.5 µm spectral range (Figure 5a). This emission band is related to the transitions ($^3$F$_2$, $^3$H$_6$) → $^3$H$_5$ (3.5–4.2 µm) and $^3$H$_5$ → $^3$H$_4$ (4.0–5.5 µm), as shown in Figure 5b. The radiative emission centered at 4 µm is clearly present but the one normally centered at 5.2 µm is not visible, mainly due to parasitic absorptions (and likely due to C-S bonds). The 3.2 µm band from the second harmonic of the pump is still visible (although attenuated by a filter) and the absorption peak at 4.2 µm is caused by CO$_2$.

The first fluorescence spectrum obtained in this double index fiber is encouraging despite the low intensity produced related to important optical losses. A new method of improvement in the synthesis of the glass and in the design of the fiber may soon be envisaged for this composition in order to considerably reduce these optical losses and to increase the intensity of their emission.

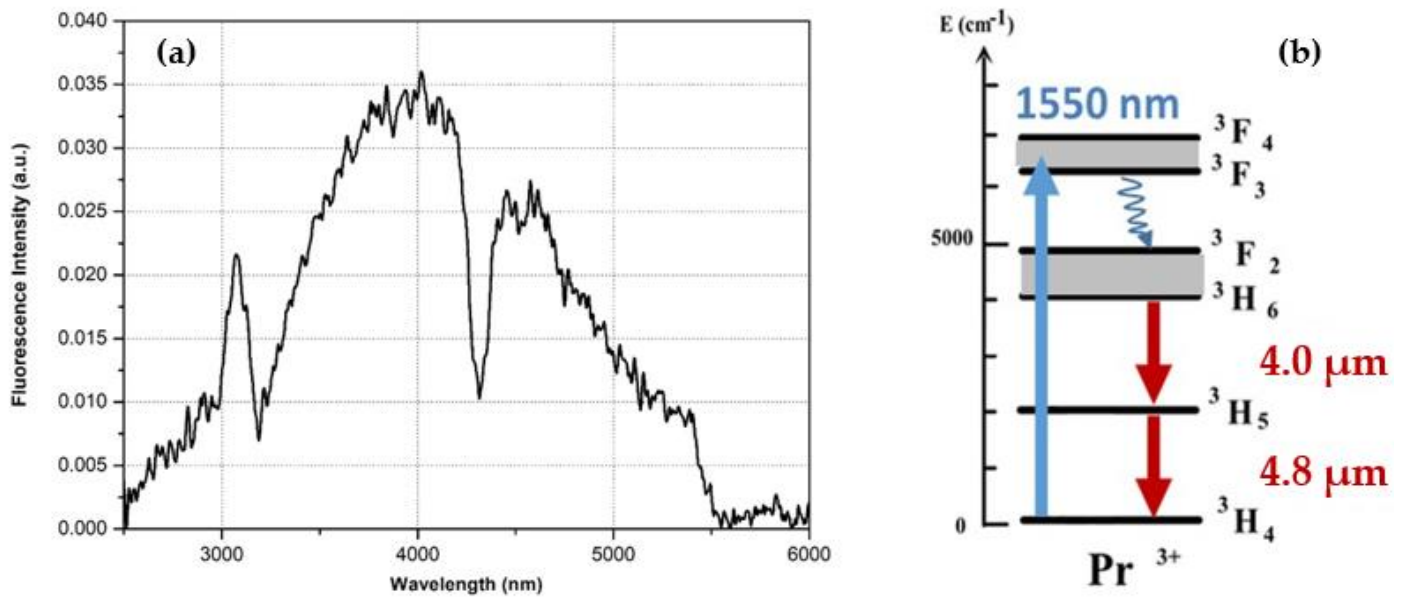

**Figure 5.** (**a**) Fluorescence intensity of the core of the double index fiber glasses ([GeS2]0.80-[Ga2S3]0.20)90-(CsCl)10 and ([GeS2]0.80-[Ga2S3]0.20)90-(CdI2)10:1000 ppmw Pr3+) as a function of the wavelength; (**b**) energy level diagram of ion Pr3+.

## 4. Conclusions

The synthesis and the characterizations of glasses belonging to the series ($[GeS_2]_{0.80}$-$[Ga_2S_3]_{0.20})_{100-x}$-$(CsCl)_x$ and ($[GeS_2]_{0.80}$-$[Ga_2S_3]_{0.20})_{100-x}$-$(CdI_2)_x$ have been implemented. Thermal properties have been studied and the photo-writability of the chosen core and clad compositions by femtosecond laser have been demonstrated. As a first step toward an MIR fiber laser, a step index fiber with a photo-writable core doped with $Pr^{3+}$ ions has been elaborated. The $Pr^{3+}$-doped double index fiber thus produced has allowed a MIR radiation around 4 µm. At this stage, even if the results are very promising, some improvements in the synthesis and fibering process will be necessary to reduce the optical losses and then to increase the level of emission before considering a fiber laser effect.

**Author Contributions:** Conceptualization, J.C., J.T., and D.L.C.; methodology, J.C., C.B.-P., F.S., and A.B.; formal analysis, P.M., and V.N.; investigation, J.C., F.S., P.M., and V.N.; writing—original draft preparation, J.C.; writing—review and editing, D.L.C., V.N., and P.C.; funding acquisition, J.T., and D.L.C. All authors have read and agreed to the published version of the manuscript.

**Funding:** This research was funded by *Agence Nationale de la Recherche* (ANR, France), grant number ANR-17-CE24-0002-02 corresponding to the COMI project.

**Institutional Review Board Statement:** Not applicable.

**Informed Consent Statement:** Not applicable.

**Data Availability Statement:** The data that support the plots within this paper are available from the corresponding author on request.

**Acknowledgments:** Julie Carcreff acknowledges both the French *Agence de l'Innovation pour la Défense* (AID) and the *Agence Nationale de la Recherche* (ANR, France) for the PhD financial supports. Platform Spectroscopy Infrared and Raman (SIR – ScanMAT, Université de Rennes 1) is acknowledged for performed Raman measurements.

**Conflicts of Interest:** The authors declare no conflict of interest related to this article.